\documentclass[11pt]{article}

\usepackage[margin=1.15in]{geometry}
\usepackage[T1]{fontenc}
\usepackage[utf8]{inputenc}
\usepackage{microtype}
\usepackage{graphicx}
\usepackage{booktabs}
\usepackage{multirow}
\usepackage{amsmath}
\usepackage{amssymb}
\usepackage{caption}
\usepackage{xcolor}
\usepackage{enumitem}
\usepackage{natbib}
\usepackage[hidelinks]{hyperref}

\setlist{itemsep=2pt,topsep=3pt,parsep=0pt}

\newcommand{\gc}[3]{#1\%~[#2,~#3]}
\newcommand{\arm}[1]{\texttt{#1}}

\title{\textbf{SPRING-5}\\[2pt]
\large Measuring Tool Documentation Repair for LLM Agents by Efficiency}

\author{Guan Zheng Huang\\
\normalsize Railtown AI\\
\normalsize \texttt{guan@railtown.ai}}

\date{\today}

\begin{document}
\maketitle

\begin{abstract}
Methods that repair tool documentation automatically are mostly evaluated on task accuracy.
We report a benchmark that scores them on cost. \textsc{Spring-5} is 20 tool families and
344 tasks, each with a deliberately degraded specification, a hand-written honest one, a
deterministic backend and an answer key, scored on calls charged against the ideal call
path. A wrong answer is charged the full call budget, so this cost index largely tracks the
completion rate, and we report calls per correct task beside it, which carries no such
charge. Over 2{,}644 baseline sessions, 54\% of wrong answers arrive with no rejected call,
no refusal and no retry. The benchmark carries three validity instruments.
A build-time ceiling assertion requires the honest specification to reach
the ideal path and fails systematically on only one family. A dud rule names the families a stronger
model solves cold and catches four. A directly measured noise floor shows that two
evaluations of a byte-identical specification differ by a 95\% factor of 0.64 to 1.56, so
about one comparison in eight crosses a conventional win threshold on noise alone, and the
cheap way of estimating that floor is biased by which families happen to go unedited.
Against these instruments, a schema-only method closes 1\% of the baseline-to-ideal cost
gap, an online self-evolution loop 13\%, one batch rewrite over every training transcript
33\%, and a variant given correctness labels 58\%. The batch advantage over the loop is the
evidence it reads: held to the loop's exposure it is no longer detectable. With five and ten tools in context the effect survives, the baseline cost does
not change, and only machine-written descriptions sent calls to the wrong tool. On BFCL, 62\% of held-out queries are already answered from the shipped
documentation, which explains a pooled null there. Code, data and runs are
released.
\end{abstract}

\section{Introduction}

A tool specification is the text an agent reads before it calls a function: what the tool
does, what each parameter means, which values are legal, what the result represents. It is
written once, by a human, usually when the endpoint is built, and then never updated from
evidence even though the agent produces a transcript of every confusion it ever has.

A growing literature repairs it automatically \citep{draft,easytool,play2prompt,docschisel}.
That work is evaluated mostly on task success. We argue this is the
wrong axis, and the reason is visible in the failure mode. When documentation is unclear an
agent does not usually fail outright. It makes three calls where one would do and then
answers correctly, or it makes one clean call, receives a well-formed number that means
something other than what was asked, and reports it with no error anywhere in the
transcript. The first costs money on every request forever and is invisible to an accuracy
metric. The second is scored as a single wrong answer, which understates a defect that will
produce a wrong answer every time it is hit.

\textsc{Spring-5} measures the first with calls per correct task, and instruments the second
with a per-session census of silent failure. Its headline cost index charges every wrong
answer the full call budget, so on its own it mostly counts lost completions, and we report
it beside both.

\paragraph{The measure, in one line.} Every family declares an \emph{ideal call path}: the
number of tool calls a correct solver needs on its held-out tasks. An arm's \textbf{cost
index} is the calls it was charged divided by that number, so 1 is perfect. The baselines
here sit between 13 and 14, and most of that is the full budget charged for each wrong
answer. \textbf{Gap closed} is the share of the distance from the
baseline's cost index down to 1 that an arm recovered, which is the figure quoted
throughout. Section~\ref{sec:cost} states both formally, together with the charging rule
for a task the agent gets wrong and the third measure, calls per correct task.

\paragraph{A benchmark of this kind has a shelf life, and should say so.} The first thing
we had to apply to our own dataset is a rule we would hand any author of another:
\emph{a family a stronger model solves from a cold start is no longer measuring
documentation, and must be named rather than averaged in}. Four of our twenty fail it on
the stronger of the two models we run, while sitting between 4.0 and 18.3 times the ideal
path on the weaker. Documentation defects do not leave production systems as models
improve; they leave our ability to measure them, and a pooled number that quietly carries
four saturated families is diluted by an amount its reader cannot recover.

\emph{The larger part of this paper is the instrumentation rather than the dataset.} A
benchmark that reports an effect size owes its reader the resolution behind it, and the
three checks of Section~\ref{sec:instruments} are what we built to supply it: what the
dataset can reach at best, how large a difference has to be before it is not noise, and
which of its families have stopped discriminating. The need is not hypothetical. A recent
audit of tool-calling benchmarks \citep{validity} finds evaluators disagreeing with expert
reviewers on 18.5\% of reviewed tasks and reports a 57.9--76.8\% spread across 23 runs of a
single benchmark, so an improvement of a
few points reported without a measured noise floor is not interpretable. We measure ours,
find it wide, and find that the cheap way of measuring it is biased. \emph{The ceiling
assertion and the noise floor have each changed a conclusion of ours}, and the dud rule
changes how any pooled number should be read, which is why we recommend all three.

\paragraph{Contributions.}
\begin{enumerate}[leftmargin=1.4em]
\item \textbf{A cost-scored benchmark} for tool-documentation repair: 20 families, 344
tasks, one planted defect per family, an honest specification and an answer key, with the
metric defined against the ideal call path (Section~\ref{sec:dataset}).
\item \textbf{Three validity instruments} that we argue any benchmark of this kind should
carry, and which we recommend independently of our dataset: a build-time ceiling assertion,
a directly measured same-spec noise floor, and a dud rule (Section~\ref{sec:instruments}).
\item \textbf{The finding that the cheap noise estimator is biased by its own coverage}, in
a direction that makes results look cleaner, and a direct estimator that costs nothing extra
(Section~\ref{sec:noise}).
\item \textbf{Silent failure measured rather than designed}: 54\% of wrong answers carry no
signal any structural rule could read, and the families that behave that way are not the
families built to (Section~\ref{sec:silence}).
\item \textbf{Five methods evaluated on it}, including the result that a single batch call
over all transcripts beats both published methods we reimplement, the finding that its lead
over the online loop comes from the evidence it reads, and the demonstration that the
cost index and calls-per-correct-task rank them differently (Section~\ref{sec:methods}).
\item \textbf{The effect measured against catalogue size}, with the findings that the
baseline cost does not change with catalogue size and that a repaired description can pull
calls onto a neighbouring tool while the shipped and hand-written texts never do
(Section~\ref{sec:multitool}).
\item \textbf{Two third-party ports and the partitioning rule they motivate}: one executes
but has no documentation headroom, the other has headroom but does not execute, and on the
second a pooled null resolves into 62\% of queries already answered from the shipped text
plus a hard remainder that is reachable in principle (Section~\ref{sec:external}).
\end{enumerate}

\section{Related Work}

\paragraph{Benchmarks for tool use.} BFCL \citep{bfcl}, StableToolBench
\citep{stabletoolbench}, API-Bank \citep{apibank}, WorkBench \citep{workbench} and
$\tau$-bench \citep{taubench} evaluate whether a model calls the right function with the
right arguments. Documentation is held fixed and treated as given, although
\citet{hsieh} show that documentation alone can carry tool use. \citet{agentsmatter} argue
that agent evaluations should report cost beside accuracy, which is the position this
benchmark takes for documentation repair. Trace-Free+ \citep{tracefree} is the closest to our concern: it
categorises interface patterns and reports that fewer than 12\% of real tool documents state
parameter constraints or selection scope, which is the gap our families are built around.

\paragraph{Documentation repair.} DRAFT \citep{draft}, EASYTOOL \citep{easytool} and
PLAY2PROMPT \citep{play2prompt} rewrite tool descriptions from feedback or self-play;
DocsChisel \citep{docschisel} diagnoses defects from failed execution traces on queries from
WorkBench and API-Bank, with a held-out split, and reports large gains over the original
documentation. Reported gains in this line shrink as baselines strengthen, and in the
comparison of \citet{tracefree} EASYTOOL falls below the original documentation on
single-step StableToolBench, so documentation edits are not safe by default. All of these
are scored on task success.

\paragraph{Harness effects.} Harness-Bench \citep{harnessbench} measures the system layer
around the model and reports a 23.8-point spread in aggregate score between harnesses,
averaged over eight model backends, while treating each harness's tool interface as native
rather than as a controlled factor. It establishes
that the layer matters without isolating this term of it.

\paragraph{Benchmark validity.} \citet{validity} is the direct motivation for
Section~\ref{sec:instruments}. Our contribution there is narrower and more actionable: three
checks that a benchmark author can run on their own data, and a demonstration that the cheap
way of computing one of them is biased.

\section{SPRING-5}\label{sec:dataset}

\subsection{One family}

\begin{figure}[t]
\centering
\includegraphics[width=\textwidth]{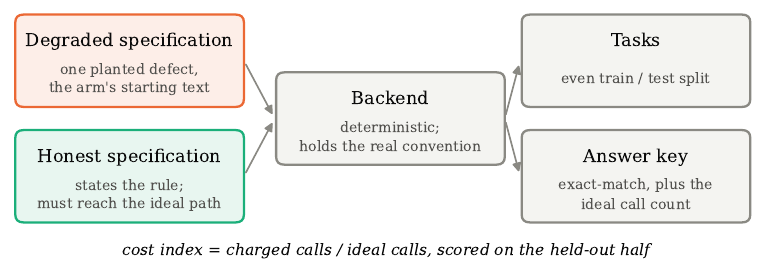}
\caption{One \textsc{Spring-5} family. Methods start from the degraded specification and are
scored on the held-out tasks. The honest specification is never given to a method; it exists
to assert that the defect is repairable in principle, and that assertion is checked at build
time.}
\label{fig:family}
\end{figure}

A family is one tool. Figure~\ref{fig:family} shows what it contains: a deterministic
backend holding the real convention, a \emph{degraded} specification carrying exactly one
planted defect, an \emph{honest} specification that states the rule, an even train/test split
of tasks, and an answer key giving both the expected value and the ideal number of calls.

The parameter set is frozen for every method: names, arity and the property set cannot
change. A proposer that invented a field would produce a specification that lies about the
function signature behind it, and freezing the set makes every method compete on the same
action space rather than on how much of the interface it is allowed to rewrite.

\subsection{Selecting the twenty}\label{sec:selection}

The twenty were filtered from a hundred candidates on four criteria, and the first two are
the ones that matter.

\textbf{Rarity.} A rule is only worth documenting if a competent engineer would not guess
it. An early candidate planted a mandatory date window and a both-ends-together pair rule,
and the baseline reached the ideal path anyway, because the model's prior over API design
already encodes those. Every family was scored on whether the convention is genuinely
unusual.

\textbf{Silence.} A confident wrong answer outranks a rejection. A defect that announces
itself through a backend error is repaired by the error message; a defect that returns a
well-formed wrong number is not. Section~\ref{sec:silence} reports how far this intent
actually carried.

\textbf{Separability and spread} complete the filter: sixteen families spread over seven
semantic groups (search ranking, time, identity, partial failure, aggregate statistics,
access policy and units) with at most three per group, the remaining four forming the two
matched pairs below, and no family whose defect could be confused with another's. The resulting defect classes are
9 \emph{output semantics}, 5 \emph{cross-parameter dependency}, 4 \emph{value convention} and
2 \emph{operational constraint}.

Four of the twenty are two matched pairs, built to isolate mechanisms rather than to
represent a class. \arm{regional\_index} and \arm{capability\_probe} are identical but for
whether the rejection message names a way forward, which on \emph{luna} separates a baseline
cost index of 25.0$\times$ from 11.6$\times$. The second pair, \arm{aggregate\_report} and
\arm{legacy\_export}, isolates a refusal made without calling the tool.

\subsection{Where the defect classes come from}\label{sec:defectsource}

The defects are planted, so the question is whether they are the defects real documentation
has. We did not invent the taxonomy. Two empirical studies of shipped API documentation
supply most of it, and we measure unplanted instances of the pattern in two third-party
benchmarks.

\textbf{The classes follow two studies of real documentation.}
\citet{apidocfails} surveyed 323 IBM software professionals and catalogued ten
documentation problems, finding that the pressing ones are about \emph{content} rather than
presentation: incompleteness and ambiguity together account for 36 problem assignments among
79 example comments, against 25 for every presentation problem combined. \textsc{Spring-5}
degrades content and never formatting, which is that finding applied.
\citet{paramrules} analysed 5{,}334 parameters that carry a rule, across nine libraries and
14{,}392 source files, and found that \textbf{86.2\% have the rule only in code, against
10.3\% only in documentation and 3.5\% in both}. A constraint enforced by the backend and
absent from the text is therefore not an unusual case to construct; it is the ordinary one.
Their category counts map onto ours directly: 18.5\% of rule sentences constrain a
parameter's \emph{format} (``\texttt{csvKeyValueDelimiter} must be exactly 1 character''),
which is our \emph{value convention}; 5.3\% describe a \emph{relation between parameters}
(``\texttt{polyLats} and \texttt{polyLons} must be equal length''), which is our
\emph{cross-parameter dependency}; and null, range and legal-value rules account for 69.1\%,
which is our \emph{operational constraint}. Our largest class, \emph{output semantics} at
nine families, has no counterpart in that study, which covers inputs only; we place it with
the content problems of \citet{apidocfails}, which are not restricted to inputs.

\textbf{The pattern occurs unplanted in documentation somebody shipped.} Two examples, both
measured rather than cited.

\emph{Value convention.} The order-lookup tool in $\tau$-bench's retail environment ships a
description whose author wrote, unprompted, ``Be careful there is a '\#' symbol at the
beginning of the order id.'' That is a maintainer adding exactly the
clarification a \textsc{Spring-5} value-convention defect removes. Deleting that sentence and
its example, and nothing else, takes the family from the ideal call path to
\textbf{17.4$\times$} it. The convention was load-bearing, and the maintainer knew it.

\emph{A code-only constraint, in its strongest form.} BFCL's \texttt{add\_postgres\_server} is scored
against an answer key that requires a \texttt{nickname} argument, and the published schema
\textbf{does not declare that parameter at all}. The user's prompt supplies the value in
seven of the eight held-out queries. Neither model we run answers a single one of them from
the shipped schema. This is the 86.2\% figure made concrete in a benchmark built by others,
and Section~\ref{sec:external} returns to it, because no method in this paper can repair it:
the parameter set is frozen for every arm, so the call the key wants cannot be emitted.

\textbf{What this does and does not establish.} It establishes that the defect classes are
drawn from measured properties of real documentation rather than from our intuition about
what would be hard, and that instances of the pattern survive in third-party artifacts. It does not
establish that our particular twenty are representative in proportion; the class mix here is
chosen for separability and spread, and Section~\ref{sec:selection} sets out that filter.
Effect sizes measured on planted defects remain effect sizes on planted defects.

\subsection{How the dataset was built and checked}\label{sec:build}

Every family was written by hand rather than generated. The degraded text, the honest text,
the backend and the answer key were authored per family, and the defect was chosen from the
classes in Section~\ref{sec:defectsource} before the text was written rather than found
afterwards.

Three checks ran on top of that. The \textbf{ceiling assertion} of
Section~\ref{sec:ceiling} is automatic and runs at build time on every family. Every answer
key is \textbf{recomputed from the backend} rather than stored by hand, so a task whose
expected value drifts from its own data fails the build. And every family, its defect and
its honest text were \textbf{reviewed by two working engineers} besides the dataset's author,
who went through all twenty with the author to check that each defect is one a competent
engineer would not guess from the interface and that each honest text states the rule
without giving away the task. That review is why the rarity criterion above is a judgement
we are willing to defend: it is not one person's intuition about what is hard.

\paragraph{Release history.} Three families, \arm{event\_count}, \arm{latency\_percentile}
and \arm{meter\_read}, were expanded from 16 tasks to 24 using parameter combinations the
backend already supported, and every answer recorded before the expansion is byte-identical
afterwards. Five others cannot be expanded without new backend data that would move recorded
answers. Runs made before the expansion use 320 tasks and three evaluation passes; later runs
use the 344 tasks described here and two. Every per-run quantity is computed at that run's
own pass count, no comparison sets a run on one release against a run on the other, and the
noise estimate of Section~\ref{sec:noise} pools passes only between evaluations over the
same task set.

The complete writer, judge and agent prompts, the schema contract every method writes
through, and the per-family degraded and honest texts are in the released artifact rather
than paraphrased here; the Reproducibility section at the end says where.

\subsection{Cost, not accuracy}\label{sec:cost}

Write $s$ for one evaluation session, $\mathrm{calls}(s)$ for the tool calls it made,
$\mathrm{ok}(s)$ for whether its final answer was correct, $B = 25$ for the call budget, and
$\iota_f$ for family $f$'s declared ideal call path summed over its held-out tasks. Over a
set of families $F$ evaluated at $R$ passes, an arm $A$ is charged
\begin{equation}
\mathrm{Charged}(A) \;=\; \sum_{s \in A}
  \begin{cases} \mathrm{calls}(s) & \mathrm{ok}(s) \\ B & \text{otherwise,} \end{cases}
\qquad
\mathrm{CostIndex}(A) \;=\; \frac{\mathrm{Charged}(A)}{R \sum_{f \in F} \iota_f} .
\label{eq:cost}
\end{equation}
\textbf{Gap closed} is the share of the available repair an arm recovered, against the
baseline serving the degraded text and the ideal path as the floor:
\begin{equation}
\mathrm{GapClosed}(A) \;=\;
\frac{\mathrm{CostIndex}(\text{baseline}) - \mathrm{CostIndex}(A)}
     {\mathrm{CostIndex}(\text{baseline}) - 1} .
\label{eq:gap}
\end{equation}
\textbf{Matched cost} drops the charging rule entirely. Let $M$ be the tasks every arm in
the comparison answered correctly on at least one pass; for each arm, average its calls
over only the passes on which it was right, then average that over $M$:
\begin{equation}
\mathrm{Matched}(A) \;=\; \frac{1}{|M|} \sum_{t \in M}
  \frac{\sum_{s \in A(t),\, \mathrm{ok}(s)} \mathrm{calls}(s)}
       {|\{s \in A(t) : \mathrm{ok}(s)\}|} ,
\label{eq:matched}
\end{equation}
and report it through Equation~\ref{eq:gap} with $\mathrm{Matched}$ in place of
$\mathrm{CostIndex}$ and the honest specification's figure in place of 1.

\paragraph{What a wrong answer costs, and why a flat cap.} Equation~\ref{eq:cost} charges
the full budget for \emph{any} session that did not end correct, which includes a session
that terminated early with a confident wrong answer as well as one that exhausted the
budget. That is a deliberate choice, and it makes the index mostly a completion measure,
so the alternatives are worth stating. Charging only the calls
actually made rewards giving up: an arm that abandons an expensive task after two calls
outscores one that solves it in six. Dropping incorrect sessions conditions the measure on
the outcome, which is the selection effect that makes a method look expensive precisely for
finishing the hard tasks. Charging an unbounded amount is not defined. The budget is a design
choice: it exceeds every correct path in the dataset, and the run is capped there regardless,
so a session is never charged more than it could have spent. The level of the index scales
with it; gap closed depends on it less, because the same charge enters the baseline and the
arm.

\emph{We do not claim the resulting index is a pure efficiency measure, and we report the
quantity that says how far it is not.} The cap term is 93--97\% of every charged total, and
the change from baseline in cost index and in completion rate disagree in sign on none of
the twenty families:
\emph{the cost index is largely the completion rate on another scale}, which is a criticism
the reader should be able to make from the paper rather than against it. This is why
\textbf{matched cost} is reported beside it everywhere and not as an appendix curiosity.
Equation~\ref{eq:matched} contains no cap term and no selection on the outcome, it is the
axis on which an arm is paid only for being cheap on work it actually did, and
Section~\ref{sec:methods} shows that the two axes rank the methods differently. A result
that appears on one and not the other is reported as such.

Intervals throughout are a \textbf{cluster bootstrap over families}, since one specification
is served per family and every task in that family is evaluated against it; resampling tasks
would understate the interval by roughly the family size.

\paragraph{Arm names.} Two arms are not methods but the ends of the scale every method is
measured between. \arm{baseline} serves the degraded specification and never edits it;
\arm{optimal} serves the hand-written honest specification and likewise never edits it, so
it is the honest text's own score and the quantity the ceiling assertion of
Section~\ref{sec:ceiling} checks. Neither reads a transcript.

The methods are named for what they do with evidence. \arm{session-evo} is the online loop,
and a \emph{ladder} is that loop run once under each of several triggers:
one session at a time, a writer woken by a trigger, and a paired trial, in which incumbent and
candidate text run on the same held-back tasks, before anything is served. The trigger
decides which sessions the writer reviews: \arm{evidence} fires on an error or a retry,
\arm{audit} on a refusal or a low score from an LLM judge that reads the session without
the answer key, and \arm{schedule} on every session. \arm{global-observer} reads every
training transcript and writes once.
\arm{labelled-observer} is the same with correctness labels and repeated rounds.
\arm{easytool} rewrites from the declared schema alone and \arm{draft} probes the tool and
rewrites iteratively; both are reimplementations of published methods on our write
channel, the one contract through which every method proposes new text. The \emph{writer}
is the model call that proposes that text.
A suffix after a slash names the trigger the online loop ran under, so
\arm{session-evo/audit} is that loop woken by the judge and \arm{session-evo/evidence} is
the same loop woken by the error-or-retry rule; a suffix after a hyphen names a variation of
a batch arm, so \arm{global-observer-k5} is the batch proposer restricted to five
transcripts per family. \emph{A row naming only a trigger is always the online loop under that
trigger.} The released runs use internal identifiers: \texttt{self-evo-ab2} is
\arm{session-evo/evidence} (written \arm{session-evo} where no other trigger is in play),
\texttt{self-evo-ab2-audit} is \arm{session-evo/audit}, \texttt{observer} is
\arm{global-observer}, \texttt{observer-k1}, \texttt{observer-k3} and
\texttt{observer-matched} are \arm{global-observer-k1}, \arm{-k3} and \arm{-k5},
\texttt{observer-ab} is \arm{global-observer-ab}, and \texttt{toolobserver} is
\arm{labelled-observer}.

\paragraph{How intervals are written.} Two quantities appear as intervals and they are not
the same thing. A \emph{gap closed} is a share and carries a percent sign on every number,
as in 12.9\%~[-2.8\%, 32.6\%]. A \emph{contrast} between two arms is a difference in points
and carries \texttt{pp} on the estimate and no percent signs inside the bracket, as in
$+23.3$pp~[$+6.8$, $+43.7$]. The distinction is deliberate: a share and a difference of
shares are different units.

\paragraph{Models.} Two models from OpenAI's GPT-5.6 family, generally available from 9 July
2026, under a 25-call budget: \texttt{gpt-5.6-luna} at low reasoning effort and
\texttt{gpt-5.6-sol} at medium, referred to throughout as \emph{luna} and \emph{sol}. Both
are called through the OpenAI Responses API with \texttt{store=False}; neither is
fine-tuned, and the dataset was written in September 2026, after both were released. The API rejects an explicit
temperature, so every number is a draw from the model's own sampling and single-run figures
are reported with intervals rather than as points.

\paragraph{Which ceiling the numbers are against.} Gap closed is computed against each
family's \emph{declared} ideal call path, the count a correct solver needs. One family,
\arm{event\_count}, does not reach its declared ceiling under \emph{luna}: its honest text
advertises a confirming call the reference solver does not require, so the family sits at
1.46$\times$ the declared ideal in every run with no unfinished sessions. It is the only
family where the declared and the measured ceiling differ systematically
(Table~\ref{tab:ceiling}), so we report against the declared one throughout.

\section{Three Instruments for Validity}\label{sec:instruments}

These are the component of this work we most expect to be reused. Each answers a question
a reader should be able to ask of any benchmark in this area, and none of the three can be
answered from a leaderboard: what is the best score this dataset admits, how large does a
difference have to be before it is not noise, and which families have stopped telling
methods apart. None is specific to our dataset and all three are inexpensive to compute.

\subsection{The ceiling assertion}\label{sec:ceiling}

\emph{The honest specification must reach the declared ideal call path.} If it does not, the
family is not measuring documentation: it is measuring something the dataset author did not
intend, and any method's failure on it will be misread as a method failure.

Checked across all six \emph{luna} runs and all four \emph{sol} runs at each run's own pass
count, six families miss on \emph{luna} and two on \emph{sol} (Table~\ref{tab:ceiling}).
All but one are explained.

\begin{table}[t]
\centering
\small
\begin{tabular}{@{}llrrr@{}}
\toprule
\textbf{Model} & \textbf{Family} & \textbf{Runs missing} & \textbf{Worst} & \textbf{Max unfinished} \\
\midrule
\multirow{6}{*}{\emph{luna}}
 & \arm{event\_count} & \textbf{6 of 6} & 1.46$\times$ & \textbf{0} \\
 & \arm{effective\_access} & 5 of 6 & 3.00$\times$ & 2 \\
 & \arm{order\_lookup} & 3 of 6 & 2.00$\times$ & 1 \\
 & \arm{fx\_rate} & 3 of 6 & 2.00$\times$ & 1 \\
 & \arm{relevance\_search} & 1 of 6 & 1.17$\times$ & 0 \\
 & \arm{legacy\_export} & 1 of 6 & 1.07$\times$ & 0 \\
\midrule
\multirow{2}{*}{\emph{sol}}
 & \arm{legacy\_export} & 1 of 4 & 20.20$\times$ & 4 \\
 & \arm{fx\_rate} & 1 of 4 & 4.00$\times$ & 2 \\
\bottomrule
\end{tabular}
\caption{Families where the honest specification misses its declared ideal path.
\arm{event\_count} is the only systematic violation: every run, and never with an unfinished
session. Every other entry carries capped sessions on a small-\emph{ideal} family, where one
unfinished task moves the ratio further than any specification could.}
\label{tab:ceiling}
\end{table}

\paragraph{The instrument has since excluded a dataset.} Before running a replication on
\textsc{Spring-2}, an earlier benchmark of ours that adapts four public APIs, we ran this
check, and it failed: with perfect documentation the agent reached only 343\% of the ideal
path and completed 51 of 60 tasks, and an online-loop arm closed 58\% of the baseline-to-ideal distance against the honest specification's 37\%. A
dataset whose declared ceiling is beaten by the methods under test cannot measure how much of
the available repair a method recovered, and no ladder was run on it. The check cost a
baseline and an \arm{optimal} arm. \emph{That is the argument for carrying the
assertion}: it is inexpensive, it runs before any claim is made, and it identified an
instrument from which we would otherwise have reported a result.

\arm{event\_count} is a genuine dataset limitation and we disclose it: its honest text
advertises a confirming call the reference solver does not need, the weaker model makes it
about half the time, and the family therefore cannot reach 100\% by construction. Zero
unfinished sessions across six runs of six is the signature of a specification property
rather than a sampling accident.

Every other entry has unfinished sessions, and a 20$\times$ violation from four capped
sessions is not a dataset defect but the cost metric's known behaviour on a small family.
\emph{The instrument is useful precisely because it separates these two cases}, and a
benchmark that reports a single pass without the unfinished count cannot.

\subsection{Evaluation noise, and a biased estimator of it}\label{sec:noise}

Every contrast in this area is a ratio between two evaluations of a tool, so how far such a
ratio moves when \emph{nothing has changed} decides what any of them mean. There are two ways
to measure it and which one you can trust depends on something you do not control.

The \textbf{opportunistic estimator} takes every arm that ends a run with a specification
byte-identical to baseline's: the ratio between their run totals is pure evaluation noise.
It costs nothing, and it is what we used first.

The \textbf{direct estimator} uses the fact that the degraded specification is fixed by the
dataset, so \emph{every baseline pass ever run is an independent draw on identical text}.
Pooling passes across runs and comparing two disjoint aggregates the size of a run total
gives the same quantity while conditioning on nothing.

\begin{table}[t]
\centering
\small
\begin{tabular}{@{}llrrrrr@{}}
\toprule
& \textbf{Estimator} & \textbf{Obs.} & \textbf{Families} & \textbf{95\% band} & \textbf{``win''} & \textbf{``harm''} \\
\midrule
\multirow{2}{*}{\emph{luna}}
 & direct (320 passes) & 188{,}538 & 20/20 & 0.642 -- 1.557 & 11.9\% & 19.0\% \\
 & opportunistic & 188 & \textbf{20/20} & 0.625 -- 1.339 & 11.7\% & 15.4\% \\
\midrule
\multirow{2}{*}{\emph{sol}}
 & direct (180 passes) & 13{,}122 & 20/20 & \textbf{0.712 -- 1.405} & \textbf{9.2\%} & \textbf{13.4\%} \\
 & opportunistic & 24 & \textbf{11/20} & 0.921 -- 1.171 & \textbf{0.0\%} & 4.2\% \\
\bottomrule
\end{tabular}
\caption{Evaluation noise on byte-identical specifications, two estimators at matched
aggregation. Both have means within 0.03 of 1.0, so neither is biased in level; they differ
in width. Passes are pooled only between evaluations over the same task set. The
opportunistic one is close to the direct one where the arms leave every family with a
control cell, a family that no arm edited, and collapses where they do not.}
\label{tab:noise}
\end{table}

\begin{figure}[t]
\centering
\includegraphics[width=\textwidth]{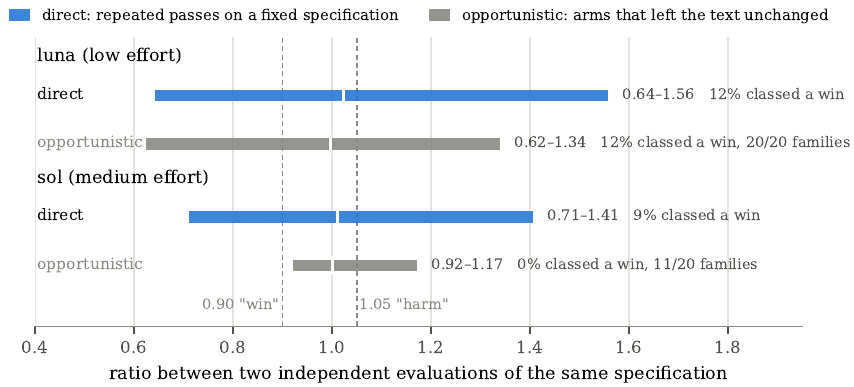}
\caption{Evaluation noise on identical text, by estimator. The opportunistic estimator tracks
the direct one on \emph{luna}, where all twenty families contribute a control cell, and
collapses on \emph{sol}, where only eleven do and those eleven are the quiet ones.}
\label{fig:noise}
\end{figure}

Table~\ref{tab:noise} and Figure~\ref{fig:noise} report both estimators at matched
aggregation.
\textbf{One comparison in eight of a byte-identical specification classifies as a win.} On
\emph{luna}, 11.9\% of same-text comparisons cross the conventional 0.90 threshold and 19.0\%
cross 1.05. Any win rate reported against fixed thresholds in this setting is inflated by
about that much and any harm rate by more.

\textbf{The cheap estimator reports a floor only for the families no arm happened to edit,
and which families those are is decided by the run rather than by the experimenter.}
On \emph{luna} all twenty families end some run unedited and the opportunistic band
is close to the direct one. On \emph{sol} only eleven ever do, and the estimator reports
\textbf{0.0\% of same-spec comparisons crossing the win threshold against a directly
measured 9.2\%}. The eleven are not a random eleven: their mean pass-to-pass standard deviation is \textbf{0.026}
against \textbf{0.134} for the nine that never yield a cell. A family yields a control cell
exactly when no method edited it, a method edits when something drew its attention, and what
draws attention is not independent of how noisy the family is. \emph{The instrument is
conditioned on the quantity it is measuring, and it fails silently in the direction that
understates variance.}

Our own first analysis used the opportunistic numbers and concluded from them that the noise
floor was a strong property of the model, \emph{sol}'s band being less than half as wide.
Measured directly the bands are 0.642--1.557 and 0.712--1.405, overlapping across most of
their range.

\textbf{The fix costs nothing.} Repeated passes over a fixed specification need no training,
no writer call and no gate, and any harness that stores per-pass results already has the
data. Our recommendation is that a benchmark in this area report a directly measured noise
floor, and that anyone measuring it opportunistically also report how many families
contributed a control cell, because that number decides whether the estimate means anything.

\subsection{The dud rule}\label{sec:duds}

\begin{figure}[t]
\centering
\includegraphics[width=0.66\textwidth]{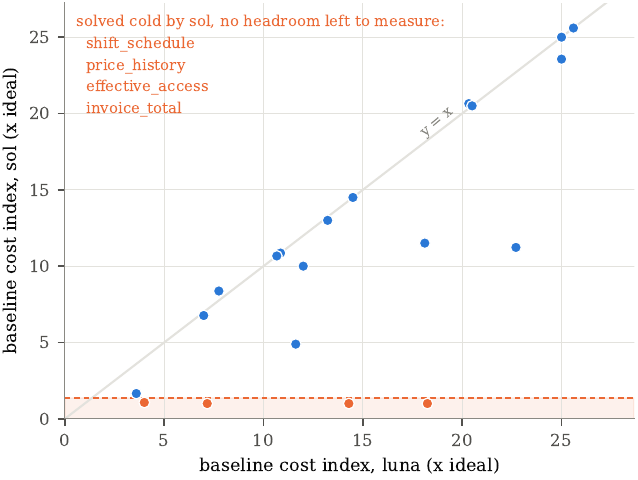}
\caption{Baseline cost index per family on both models. Sixteen families stay above the dud
threshold of 1.35$\times$ on both, the closest being \arm{account\_link} at 1.66$\times$ on \emph{sol}. Four fall to the floor on \emph{sol}: they are solved from a cold start,
and any method scored on them can only be measured against noise. This is how a benchmark of
this kind ages.}
\label{fig:invariance}
\end{figure}

The rule stated in Section~\ref{sec:dataset} is that a family a model solves from a cold
start has no headroom, so averaging it into a pooled number dilutes every other family.
This section applies it. Figure~\ref{fig:invariance} plots each family's baseline cost
index on the two models.

Four families, \arm{effective\_access}, \arm{invoice\_total}, \arm{price\_history} and
\arm{shift\_schedule}, drop below 1.35$\times$ ideal on \emph{sol} while sitting between
4.0$\times$ and 18.3$\times$ on \emph{luna}. \textbf{This is the most consequential limitation of
the instrument.} Documentation defects do not disappear from production systems as models improve;
they disappear from our ability to measure them. Any headline number in this area is
perishable, and a successor dataset will need to be harder rather than larger. We recommend
reporting the dud count alongside any result, and naming which families it covers.

\section{Results}\label{sec:results}

\subsection{Silent failure is a behaviour, not a design property}\label{sec:silence}

The families were selected partly on silence, and thirteen carry a defect in a class we
expected to fail silently. That is a statement about construction. What the sessions do is a
different question and we measured it (Table~\ref{tab:silence}): a wrong answer counts as
silent when the transcript carries nothing a structural rule could fire on, meaning no
rejected call, no refusal and no retry.

\begin{table}[t]
\centering
\small
\begin{tabular}{@{}lrr@{}}
\toprule
& \emph{luna} & \emph{sol} \\
\midrule
baseline sessions & 2644 & 664 \\
wrong & 1597 & 303 \\
\textbf{wrong and silent} & \textbf{857 (54\%)} & \textbf{153 (50\%)} \\
families majority-silent & \textbf{11 of 20} & 7 of 20 \\
families entirely silent & 8 & 6 \\
\bottomrule
\end{tabular}
\caption{Silent failure measured per session. More than half of all wrong answers carry no
signal any structural rule could read.}
\label{tab:silence}
\end{table}

\begin{figure}[t]
\centering
\includegraphics[width=\textwidth]{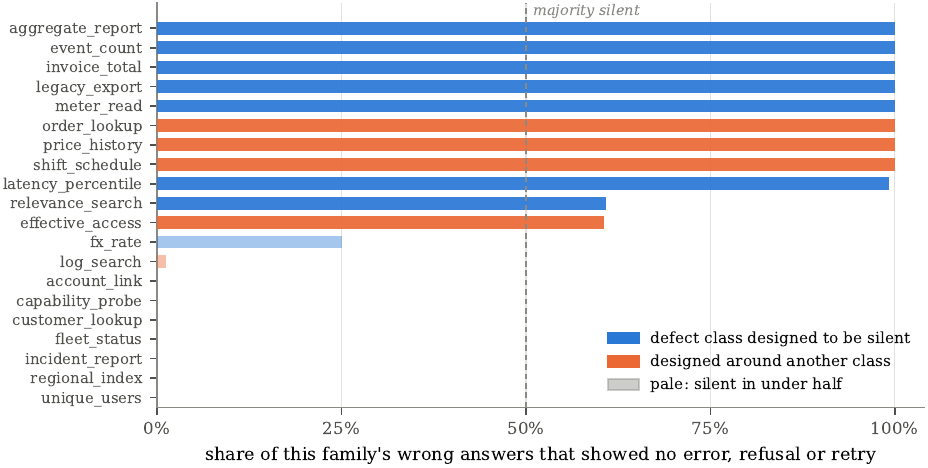}
\caption{Per family, the share of wrong answers that showed no error, refusal or retry, on
\emph{luna}. Colour is the defect class the family was \emph{designed} around. The design
does not determine the behaviour: six families intended to be silent are not, and four
intended otherwise are.}
\label{fig:silent}
\end{figure}

\textbf{Intent and behaviour agree on only seven families.} Six built to fail silently do not,
announcing themselves through rejections or refusals often enough for a structural rule to
reach them, and four built around other defect classes are silent in practice
(Figure~\ref{fig:silent}). The per-session share, not a family count, is the number worth
quoting, and it is also the number that supports the argument: \emph{more than half of all
failures here are invisible to any rule over transcript structure}. Neither an accuracy score
nor the cost index, which charges a silent wrong answer exactly as it charges a loud one,
separates these from other failures; the census does, which is why it is reported beside
both.

\subsection{Five methods}\label{sec:methods}

We evaluate two published methods, two observer architectures and one online
self-evolution loop, all reimplemented on an identical channel and action space.
\arm{easytool} rewrites from the schema alone. \arm{draft} probes the tool and rewrites
iteratively. \arm{global-observer} makes one writer call over every training transcript.
\arm{labelled-observer} is \arm{global-observer} plus ground-truth correctness labels, which no
deployment has; it prices the label rather than the method. \arm{session-evo} is an online
loop with a trigger, a writer and a paired acceptance test.

\begin{table}[t]
\centering
\small
\begin{tabular}{@{}llrrrr@{}}
\toprule
\textbf{Method} & \textbf{Sees} & \textbf{Gap closed} & \textbf{95\% CI} & \textbf{Matched} & \textbf{Tok.} \\
\midrule
\arm{easytool} & schema only & 1.0\% & [-10.3\%, 12.3\%] & 18.0\% & 12k \\
\arm{draft} & own probing & 10.4\% & [-1.4\%, 25.6\%] & 14.0\% & 146k \\
\arm{session-evo} & one per review & 12.9\% & [-2.8\%, 32.6\%] & 40.0\% & 222k \\
\textbf{\arm{global-observer}} & \textbf{all, one call} & \textbf{32.9\%} & \textbf{[14.4\%, 55.0\%]} & \textbf{56.0\%} & 156k \\
\arm{labelled-observer}$^\dagger$ & all + labels & \textbf{58.2\%} & [35.2\%, 79.5\%] & 32.0\% & 475k \\
\bottomrule
\end{tabular}
\caption{Five methods on \textsc{Spring-5}, \emph{luna}, 20 families at two passes. ``Gap
closed'' is on the cost index; ``Matched'' is gap closed on calls per correct task over the
60 tasks every method solved. Table~\ref{tab:exposure} reports separate runs, so its figures
for the same arms differ by run-to-run noise.
$\dagger$ requires a correctness label and is a ceiling, not a deployable method.}
\label{tab:methods}
\end{table}

\paragraph{Two arms are not comparable to the others, and we separate them.}
The labelled arm carries ground-truth labels. No deployment has those, so it is
not a competing method but a \emph{ceiling}: it prices what a label is worth, in the role
an oracle plays elsewhere in this paper, and its number should never be quoted alongside the
others as though they were the same kind of thing. \arm{global-observer} approaches that
boundary in a different way. It retains every training transcript until it writes,
which is a short-lived cache rather than a permanent log, but it is a stronger data-retention
assumption than the online loop makes, and in a setting where transcripts cannot be retained
at all it is simply unavailable. We report it as a method, with that condition attached.

Four results follow, with \arm{labelled-observer} read as a ceiling throughout.

\textbf{A schema-only method recovers nothing.} \arm{easytool} at 1.0\% is the direct
confirmation that a method rewriting from the schema alone cannot state a convention the
schema never contained. \arm{draft} and the online loop are indistinguishable from each other
and their intervals include zero.

\textbf{A single batch call beats both published methods.} \arm{global-observer} closes
32.9\% with no trigger, no acceptance test, no iteration and one writer call, beating
\arm{draft} by $+22.5$pp [$+6.3$, $+43.3$] and \arm{easytool} by $+31.9$pp [$+13.6$, $+53.5$] with intervals
clear of zero, at two thirds of the online loop's training tokens. The only thing it changes
is \emph{how much evidence reaches the writer at once}. We report this as a benchmark result
rather than a proposed method; its implications for loop design are outside the scope of
this paper.

\textbf{Against the online loop the advantage is consistent in direction.} In this run
\arm{global-observer} leads \arm{session-evo} by $+20.0$pp $[-5.1, +48.1]$, an interval
that includes zero. The two exposure runs of Section~\ref{sec:exposure} repeat the
comparison on \emph{luna} at $+23.3$pp $[+6.8, +43.7]$ and $+23.9$pp $[+6.0, +44.7]$, and on
\emph{sol}, over eight families chosen
for high baseline cost, \arm{global-observer} closes \gc{25.5}{0.4\%}{61.9\%} against the
loop's \gc{8.1}{-2.9\%}{27.5\%}, a contrast of $+17.4$pp $[+1.1, +45.9]$, on \emph{fewer}
training tokens (71k against 81k) and the same number of training tool calls. On calls per
correct task it leads there 79.2\% to 56.2\%.

\begin{figure}[t]
\centering
\includegraphics[width=\textwidth]{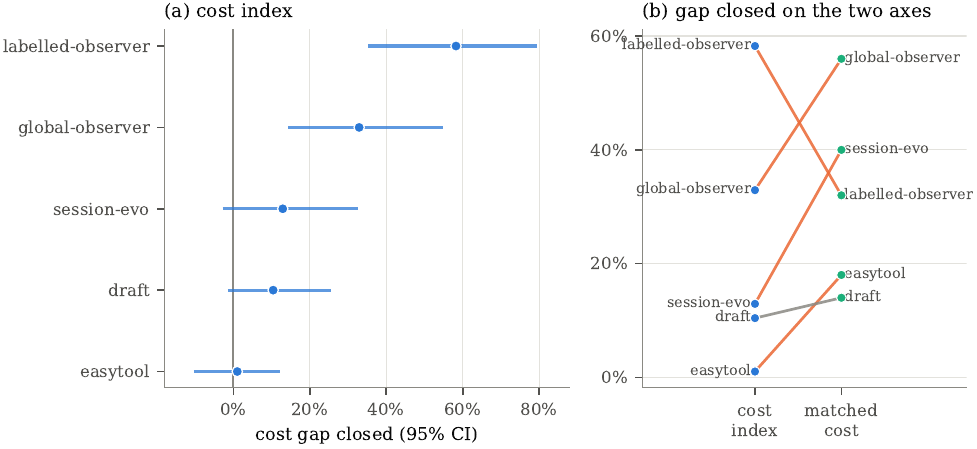}
\caption{\textbf{(a)} Cost gap closed with cluster-bootstrap intervals over families.
\textbf{(b)} The same methods on the two axes. Lines that cross are methods the axes rank
differently; orange marks a change of more than 12 points.}
\label{fig:methods}
\end{figure}

\textbf{The two axes order the methods differently, on a narrow scale.} On calls per correct
task, \arm{labelled-observer} falls from first to third and \arm{session-evo} rises from
third to second (Figure~\ref{fig:methods}b). The scale is small: over the 60 tasks every
method solved, the baseline spends 1.43 calls per correct answer and the honest text 1.02,
and the methods sit between 1.20 and 1.38, so the matched figures carry no intervals here and
the reordering should be read as a direction. A correctness label tells the writer which
families are broken; it does not make the resulting text more efficient on the families
everyone already solves. Among the methods that need no label, \arm{global-observer} leads on
both axes.

\paragraph{How much the matched set depends on which arms define it.} Equation
\ref{eq:matched} is computed on the intersection of what every arm solved, so a reader is
entitled to ask whether the 60 tasks, and the ordering on them, are an artifact of the five
methods we happened to compare. Recomputing with each method dropped in turn, the set grows
to between 60 and 65 tasks, never by more than five, and \emph{the ordering of the
remaining methods is identical in all five cases}. What moves is the level attached to arms
far from the top. Dropping \arm{session-evo} admits four tasks that only it failed and
takes \arm{draft} from 14.0\% to $-13.7$\%, because on those four \arm{draft} spends more
per correct answer than \arm{baseline} does; \arm{draft} is last under both readings.
Dropping \arm{labelled-observer} changes nothing at all, the set staying at the same 60
tasks. \emph{What the table is used for, that the two axes rank the methods differently,
does not depend on the composition of the matched set. The percentage attached to a
low-ranked arm does, and should be read as such.}

\subsection{Is the batch advantage architecture, or just more evidence?}\label{sec:exposure}

\arm{global-observer} beats the online loop, and the obvious objection is that it is not a fair
fight. The batch arm reads every training transcript in one call. The loop's writer is shown
one transcript per review and holds at most three reviews per family, so over the method run
of Table~\ref{tab:methods} it read \textbf{24} transcripts against the batch arm's
\textbf{172}. If the gap is exposure
rather than architecture, the comparison says nothing about batch versus online.

We tested it directly with batch arms that read fewer transcripts. \arm{global-observer-k1}
reads one per family, about the loop's total exposure, and \arm{global-observer-k3} reads
three, the loop's per-family maximum. A second run holds \arm{global-observer-k5}, at five
per family, and \arm{global-observer-ab}, the batch proposer with its rewrite put to a paired
trial on tasks held back from the writer.

\begin{table}[t]
\centering
\small
\begin{tabular}{@{}lrrr@{}}
\toprule
\textbf{Arm} & \textbf{Transcripts read} & \textbf{Cost index} & \textbf{Gap closed} \\
\midrule
\multicolumn{4}{@{}l}{\emph{Exposure run}} \\
\arm{baseline} & 0 & 13.35$\times$ & -- \\
\arm{optimal} & 0 & 1.20$\times$ & \gc{98.4}{96.1\%}{100.0\%} \\
\arm{session-evo} (online) & 23 & 12.04$\times$ & \gc{10.6}{2.4\%}{22.8\%} \\
\arm{global-observer-k1} & \textbf{20} & 11.59$\times$ & \gc{14.2}{3.2\%}{29.3\%} \\
\arm{global-observer-k3} & 60 & 10.28$\times$ & \gc{24.9}{8.3\%}{44.1\%} \\
\arm{global-observer} & 172 & 9.16$\times$ & \gc{34.0}{14.8\%}{56.6\%} \\
\midrule
\multicolumn{4}{@{}l}{\emph{Second run}} \\
\arm{session-evo} (online) & 23 & 12.31$\times$ & \gc{10.1}{-2.1\%}{25.8\%} \\
\arm{global-observer-k5} & 100 & 9.89$\times$ & \gc{29.3}{11.3\%}{50.8\%} \\
\arm{global-observer} & 172 & 9.31$\times$ & \gc{33.9}{14.2\%}{56.3\%} \\
\arm{global-observer-ab} & 109 & 10.37$\times$ & \gc{25.5}{8.3\%}{45.7\%} \\
\bottomrule
\end{tabular}
\caption{The batch proposer at several transcript budgets against the online loop,
\emph{luna}, 20 families, two evaluation passes, in two separate runs. Transcripts read is
what reached a writer, not what was executed; for the online loop it is one per review.}
\label{tab:exposure}
\end{table}

\textbf{The advantage is mostly the evidence} (Table~\ref{tab:exposure}). Held
to the loop's exposure, 20 transcripts against its 23, the batch arm closes 14.2\% against
the loop's 10.6\%, a contrast of $+3.6$pp $[-9.8, +16.3]$ that is consistent with no
difference. With three transcripts per family it closes 24.9\%, and with every transcript
34.0\%, $+19.7$pp $[+3.5, +39.9]$ above the one-transcript arm and $+23.3$pp
$[+6.8, +43.7]$ above the loop. \emph{Writing once over everything is not what wins;
reading more is}, although at matched exposure the interval still admits a difference of up
to 16 points. The loop read 23 of a possible 60 transcripts because its trigger fired on few
sessions, so how much it reads, rather than its incremental shape, is what separates the two
here. The second run
agrees: at five transcripts per family the batch arm closes 29.3\%, between the three-transcript
and full arms, and the full batch arm leads the loop there by $+23.9$pp $[+6.0, +44.7]$.

\emph{This is a result in a static regime.} Every transcript in a batch here was produced
against one unchanging backend, which is the condition a single reconciling rewrite needs,
and \textsc{Spring-5} supplies it by construction. Where an endpoint changes mid-window a
batch spanning the change mixes two behaviours into one rewrite and cannot tell which
transcripts describe which, while an online loop at least observes the change in order.
Section~\ref{sec:limits} states the test that would price that, and we would expect it to
cut the other way.

\textbf{The paired trial costs more than it saves here.} \arm{global-observer-ab} loses
$-8.5$pp $[-19.4, -0.7]$ against plain \arm{global-observer}, an interval clear of zero.
The paired trial has to hold back 30\% of the training half to have anything disjoint to test
on, so it buys an acceptance test by spending the evidence that made the proposer good. On
this dataset that trade is negative. We report it because a paired trial is easy to treat as
free, and on a batch proposer it is not.

\subsection{Power}\label{sec:power}

Simulating a family-level test at the between-family spread we measure, an arm-to-arm
difference in cost ratio with standard deviation 0.45 across families, twenty families
detect a true difference of 0.10 in cost ratio \textbf{19\%} of the time and one of 0.20
\textbf{54\%} of the time; 80\% power needs about 39 families at 0.20 and 162 at 0.10. We
state this because the field ranks methods at this scale as a matter of course, and because it explains
the shape of Table~\ref{tab:methods}: the contrasts that clear zero are the large ones, and
the three arms clustered between 1\% and 13\% cannot be separated by this instrument at this
size. A benchmark that cannot separate nearby methods should say so rather than publish the
ordering.

\section{Catalogue Size}\label{sec:multitool}

Every number above advertises one tool. Production agents carry many, and with more tools in
context the agent must first find the right one, so a result measured at one tool does not
transfer by assumption. We re-ran the ladder with the family under test advertised alone,
alongside four others and alongside nine, drawn by rotation through the catalogue, with every
other setting identical and on the current release. The ideal call path is unchanged by the bundle, so
gap closed stays comparable; what changes is that the agent must first find the tool.

\begin{table}[t]
\centering
\small
\begin{tabular}{@{}lrrrrr@{}}
\toprule
\textbf{Run} & \textbf{Tools} & \textbf{Baseline} & \textbf{\arm{optimal}} &
\textbf{\arm{\dots/evidence}} & \textbf{\arm{\dots/audit}} \\
\midrule
run 1     & 1  & 13.3$\times$ & 99.4\% & \gc{11.4}{1.2\%}{25.1\%} & \gc{10.8}{-3.4\%}{28.3\%} \\
run 2     & 1  & 13.8$\times$ & 98.4\% & \gc{13.6}{4.2\%}{26.2\%} & \gc{14.4}{-0.7\%}{32.1\%} \\
\midrule
run 1     & 5  & 13.3$\times$ & 99.8\% & \gc{18.5}{6.9\%}{33.8\%} & \gc{9.6}{-8.3\%}{28.9\%} \\
run 2     & 5  & 13.3$\times$ & 98.8\% & \gc{9.3}{1.2\%}{20.6\%} & \gc{7.3}{-10.6\%}{26.1\%} \\
\midrule
run 1     & 10 & 13.7$\times$ & 97.3\% & \gc{15.2}{4.2\%}{28.9\%} & \gc{5.9}{-9.4\%}{23.6\%} \\
\bottomrule
\end{tabular}
\caption{The ladder at three catalogue sizes, \emph{luna}, two evaluation passes, with a
replicate at one and five tools. The two right-hand columns are the same
\arm{session-evo} online loop of Table~\ref{tab:methods} under its two triggers,
abbreviated here to the trigger name; \arm{optimal} is the honest specification's own
score. Baseline is the cost index; the arms are gap closed with a
cluster bootstrap over families. Adding distractors does not remove the effect, and the
between-run spread at a fixed catalogue size is larger than the difference between sizes.}
\label{tab:bundles}
\end{table}

\textbf{The result survives, and the ceiling holds} (Table~\ref{tab:bundles}). With one tool
the loop closes 10.8\% to 14.4\% of the cost gap and at five and ten tools 5.9\% to 18.5\%;
the spread between two runs at one catalogue size is as large as the difference between
sizes. \arm{optimal} reaches 97.3--99.8\% at every catalogue size, so the honest
specification still reaches the declared ideal path with nine distractors present and the
ceiling assertion of Section~\ref{sec:ceiling} is not a single-tool artifact.

\textbf{The baseline cost does not move with catalogue size.} It is 13.3$\times$ and
13.8$\times$ at one tool and 13.3$\times$, 13.3$\times$ and 13.7$\times$ at five and ten.
Our families were built to be semantically distant, so a bundle adds context without adding
much selection pressure, and in this regime the cost of a defect is the same whether the
tool arrives alone or among nine others. A catalogue of near-siblings would be a harder test,
and the next subsection shows where one appears.

\subsection{A rewrite can cause misrouting; the shipped and hand-written texts never did}

The counter that matters here is calls spent on a tool other than the one under test.

\textbf{Across 2{,}064 \arm{baseline} and \arm{optimal} sessions at bundles of five and ten,
there were zero off-target calls.} The arms that served a machine-written
specification did not: 10 of 416 calls in one run, 5 of 417 in another, 11 of 388 and 1 of
391 in the third. \emph{Every one of them, in every run, came from the same family}, and all
went to the same neighbour: \arm{regional\_index}, whose served text drew calls to
\arm{capability\_probe}.

That pair is not an accident. \arm{regional\_index} and \arm{capability\_probe} are the
matched pair of Section~\ref{sec:selection}, built to be identical but for their rejection
message, and they are therefore the closest siblings the dataset contains. The mechanism is
visible in the text. The degraded description reads ``Search the regional document index for
a region.'' The writer replaced it with ``Search the regional document index. The specified
region must be entitled for the current principal; requests for unauthorized regions are
rejected by the backend.'' The added clause is accurate and it repairs the defect. It also
introduces entitlement and authorisation vocabulary that was not there before, and in a
catalogue containing a capability-shaped sibling the agent followed it to the wrong tool.

\textbf{This is a cost a single-tool benchmark cannot charge.} A rewrite is scored locally,
on whether the description now states the rule, and the paired trial of
Section~\ref{sec:methods} compares candidates on the family they belong to. Neither sees a call that went somewhere
else. The effect is small here because our families were deliberately built to be
semantically distant, and it appears only at the one place where that was relaxed; on a real
catalogue, where near-siblings are the norm rather than a constructed pair,
Section~\ref{sec:external}'s BFCL observation suggests it would be larger. \emph{A tool
description is not a local object, and a documentation-repair method that optimises one
description at a time is optimising an objective that leaves this cost out.}

\section{Two Third-Party Ports, and What They Cannot Show}\label{sec:external}

\textsc{Spring-5} exists because we could not find a public benchmark that both
\emph{executes}, so that calls can be counted, and \emph{leaves documentation headroom}, so
that there is repair to measure. This section reports the two we ported, because between
them they say why that combination is rare, and because one of them yields a partitioning
result we think anyone evaluating documentation repair should run before quoting a pooled
number. \emph{Neither is a transfer result for this paper's measure, and we do not present
either as one.}

\paragraph{The executing port has no headroom.} A port of $\tau$-bench's retail environment
runs under our harness with the cost index fully computable. With the shipped text the
agent reaches the \emph{ideal call path on seven of seven families}, so the
baseline-to-ideal distance that Equation~\ref{eq:gap} divides by is zero and no method can
be scored. Ablating the spans its maintainers wrote opens headroom on one family, which is
too few to bootstrap over. The benchmark is not deficient; it is simply documentation that
was written carefully for the model it ships against.

\paragraph{The port with headroom does not execute.} BFCL v4 \texttt{live\_multiple} is 20
real functions carrying twelve or more queries each, 637 queries, third-party schemas and a
third-party answer key. Reconstructing a call directly from the key scores 637 of 637, so
our scorer follows the benchmark's convention. But BFCL scores one emitted call against a
key by structure: nothing executes, there is no backend, no retry and no call budget, so
\textbf{the cost index cannot be computed here at all} and the comparison necessarily runs
on task accuracy, which cannot see the extra calls a correct answer costs. The
\arm{evidence} trigger cannot fire either, since with one shot and no execution there is no
error turn and no repeated call for it to read.

\emph{We therefore report BFCL for the two things it can establish}, both of which are about
the instrument rather than about our effect sizes: whether a benchmark of this sort can be
pooled before it is partitioned, and whether failures on documentation nobody degraded are
of the kind we plant. Readers wanting the cost claim tested off our own dataset will not
find it here, and Section~\ref{sec:limits} says why we do not have it.

\textbf{Pooled over every held-out query the result is a null.} Against a 59.1\% baseline the
\arm{audit} loop gains $+3.4$pp with a 95\% interval of $[-0.4, +7.4]$ and a scheduled variant
$+1.9$ $[+0.0, +3.9]$. An improvement of this size is not negligible on a public leaderboard.
It is, however, barely separable from zero here, and pooling turns out to be the wrong way to
read it.

\subsection{The benchmark is two populations}

Running the \emph{stronger} model on the \emph{unmodified} schema partitions the held-out
queries by whether the shipped documentation is sufficient for a capable caller. This is the
dud rule of Section~\ref{sec:duds} applied at query level to someone else's data.

\begin{table}[t]
\centering
\small
\begin{tabular}{@{}lrrrr@{}}
\toprule
& \textbf{Queries} & \textbf{baseline} & \textbf{\arm{audit}} & \textbf{\arm{schedule}} \\
\midrule
Pooled & 323 & 59.1\% & $+3.4$ [$-0.4$, $+7.4$] & $+1.9$ [$+0.0$, $+3.9$] \\
\midrule
Saturated & 200 & \textbf{94.5\%} & $+3.0$ [$-1.7$, $+7.4$] & $+0.0$ [$-2.5$, $+2.6$] \\
Hard & 98 & \textbf{2.0\%} & $+5.1$ [$-1.4$, $+11.7$] & \textbf{$+6.1$ [$+1.0$, $+11.2$]} \\
\bottomrule
\end{tabular}
\caption{The BFCL transfer, pooled and split by whether the stronger model answers the query
from the shipped schema. Arms are reported as the change from that stratum's own baseline,
with a cluster bootstrap over families. The two strata differ by 93 points of baseline
accuracy. The two families carrying an out-of-schema query are excluded from the split rows
for the reason given in Appendix~\ref{app:bfcl}, and a family contributes to a stratum only
where it has queries
in it, so the split rows sum to 298 over 17 and 16 families rather than to 323 over 20.}
\label{tab:bfclstrata}
\end{table}

Table~\ref{tab:bfclstrata} reports both splits. \textbf{Sixty-two percent of the benchmark
has no headroom.} On the saturated stratum the
weaker model already answers 94.5\% from the shipped text. Nothing a rewrite does can show up
there, and averaging it against the rest is what produced the null: the pooled figure is
dominated by queries on which the shipped documentation was never the binding constraint.

\textbf{The hard stratum is not model-limited in the way it first appears.} There
\arm{audit} moves the baseline from 2.0\% to 7.1\% and \arm{schedule} to 8.2\%, the
latter with an interval clear of zero. The counts are small, 2 correct against 7 and 8, so we
treat this as evidence that the stratum is reachable rather than as a measurement of how far.
What it rules out is the reading that these queries are simply beyond the model.

\subsection{What blocks the hard stratum}\label{sec:bfclcauses}

We read all 121 held-out queries that neither model answers from the shipped schema and
assigned each a cause, writing every assignment as a rule so the count reproduces. The 121
are counted over all twenty families, while the hard stratum of
Table~\ref{tab:bfclstrata} leaves out the two families carrying an out-of-schema query, which
is why the two figures differ.
Appendix~\ref{app:bfcl} gives the full taxonomy; three figures carry the argument.
\textbf{Routing is correct on 119 of the 121}, so almost none of this is tool selection, and
\textbf{on 78 of them both models emit a byte-identical call} that the key rejects, which is
two independently sampled models agreeing on a reading of the same text rather than two
models guessing. \textbf{In 72 of the 121, 59.5\%, the fix lies in the description of the
function under test}, which is the action space every method in this paper is confined to.

\textbf{One class accounts for 50 of those 72, and it is a class we plant deliberately.} The
description tells the caller to pass a sentinel the declared type will not carry:
\texttt{Music\_3\_LookupMusic.year} is declared \texttt{integer} and its description reads
``Use 'dontcare' to include songs from any year''. The advice cannot be followed through a
typed function-tool interface, and omitting the parameter, which the key also accepts in
every such case, is the one action the text never mentions. \emph{That is our value-convention
class occurring unplanted, in a third-party benchmark, at scale}, and it is the single
observation from this transfer that bears directly on the dataset's construction. It is also
the clearest case for repairing documentation offline rather than expecting the caller to
cope: the agent gets one shot with no execution and no feedback, while the contradiction is
visible in the schema before any call is made, since \texttt{year} is declared
\texttt{integer} while its own \texttt{default} field holds the string \texttt{dontcare}.

The remaining 49 are outside what any rewrite here can do: dates the request never grounds,
a key that spells one kind of entity two ways inside a single family, values the request
never states, and the undeclared parameter of Section~\ref{sec:defectsource}. We record them
because \emph{a benchmark's unreachable fraction is a property a method's score should be
reported against}, and 40.5\% is large enough to change how a null on the remainder reads.

\subsection{What we take from the transfer}

The honest summary is narrower than either a null or a positive. \emph{Neither port settles
whether this loop transfers}: one has no headroom to measure against, and on the other 62\%
of the queries are already answered from the shipped text while the part that is not
supports only a direction at these counts. We report both because the diagnosis is reusable
and costs one run of a strong model on unmodified documentation: \emph{any benchmark used
to evaluate documentation repair should be partitioned this way before a pooled number is
quoted}. Ours would fail the same check on four families (Section~\ref{sec:duds}), which is
why the dud rule is stated as an instrument rather than as a disclosure.

Two further observations from the transfer are worth recording because \textsc{Spring-5}
cannot see them. \textbf{An absolute trigger threshold did not port}: a fixed judge-score
threshold set on our dataset never fired once here, because a judge scoring a single call with no tool
response sits on a different part of the scale; a rank-based rule needs no calibration and
does fire. And \textbf{a tool description is not a local object}: BFCL ships deliberate
near-miss siblings, and rewriting one description moved routing accuracy from 100\% to 87.5\%
on one family, taking task accuracy down with it even though the new text read better in
isolation. Section~\ref{sec:multitool} measures that effect on our own dataset.

\section{Limitations}\label{sec:limits}

\begin{itemize}[leftmargin=1.4em]
\item \textbf{The defects are ours, and the cost claim has never been tested off this
dataset.} Every reported effect size is measured on a dataset whose defects we planted, and
Section~\ref{sec:external} reports the two ports that were meant to fix that and could not:
$\tau$-bench executes but leaves no headroom, BFCL leaves headroom but does not execute, so the
cost index is computable on one and meaningful on neither. \emph{That absence is the reason
this dataset exists, and it is also the largest open question about it.} The instruments,
rather than the effect sizes, are the contribution we consider best supported.
\item \textbf{Twenty families, low power.} A true difference of 0.10 in cost ratio is
detected 19\% of the time (Section~\ref{sec:power}). The benchmark does not support
separating methods whose true difference is small, and no such separation is claimed here.
\item \textbf{Four families are already saturated on the stronger model.} The instrument
ages with the models it is run against, and the families whose behaviour is stable across
both models are its durable component.
\item \textbf{Catalogue realism.} Section~\ref{sec:multitool} runs the ladder at one, five and
ten tools and the effect survives, but our families were built to be semantically distant and the
one constructed near-pair is the only place misrouting appeared. A catalogue of genuine
near-siblings would stress this harder than we can.
\item \textbf{\arm{draft} and \arm{easytool} are reimplementations} on our channel and
action space. This supports a controlled comparison of architectures; it does not constitute
a replication of the originally reported results.
\item \textbf{\arm{global-observer} assumes the transcripts can be kept.} It retains every
training transcript until it writes, and it reads them all under the degraded specification,
whereas an online loop discards each transcript after its review and edits the text it is
also being evaluated on.

\item \textbf{Batch observation assumes the tool did not change under it.} \textsc{Spring-5}
serves one version of each tool, so every transcript in an \arm{global-observer} batch was produced
against identical behaviour. This is the condition a batch method requires, and it is not
guaranteed in practice: where an endpoint changes mid-window, a batch spanning the change
mixes two behaviours into one rewrite and cannot tell which transcripts describe which, while
an online loop at least observes the change in order. We verified that no transcript in any
arm came from the held-out split, so the comparison reported here is sound. \emph{On a
benchmark with version drift the batch advantage could narrow or reverse}, which is the
natural next test.
\end{itemize}

\section{Conclusion}

We built a benchmark for tool-documentation repair scored on cost as well as accuracy, with
calls per correct task beside a cost index that charges every wrong answer the full budget,
and with a census of the more than half of its failures that produce no error, no refusal
and no retry. The dataset is 20 families and
344 tasks and we release it, but we expect the three validity instruments to outlast it: a
build-time assertion that the honest specification reaches the ideal path, a dud rule for
families a stronger model solves cold, and a directly measured noise floor.

The last of those produced the result we least expected. Repeated evaluations of a
byte-identical specification differ by a 95\% factor of 0.64 to 1.56, so roughly one
comparison in eight crosses a conventional win threshold on noise alone; and the cheap way of
estimating that floor is biased by its own coverage, reporting zero same-spec wins on one of
our models where the directly measured figure is about one in eleven. Our own first analysis
used the biased version. Anyone reporting a win rate in this area without a directly measured floor is
reporting an unknown mixture of effect and variance.

\section*{Reproducibility}

Dataset, harness, every run directory, the figure-generating scripts and the JSON file
backing every number are released at \url{https://github.com/CoronRing/DATE}. Each figure and table is generated from that file rather
than transcribed.

\textbf{Prompts and contracts are released verbatim, not paraphrased.} The agent system
prompt, the writer prompt each method is given, the judge prompt and trust scale, and the
schema contract that fixes what a method may rewrite are in the harness source at the paths
below, so a reader can check that the methods were held to the same action space rather than
take our word for it.

\begin{itemize}[leftmargin=1.4em]
\item \arm{arena/agent.py}: the agent contract, the answer check, the writer contract
that freezes the parameter set, and the judge prompt with its trust scale.
\item \arm{arena/methods.py}: one function per method, each carrying the deviations from
its reference implementation in its own docstring.
\item \arm{arena/spring5/}: the twenty families, giving degraded text, honest text,
backend and answer key, one module per semantic group.
\item \arm{arena/metrics.py}: the cost index, gap closed, the cluster bootstrap, the
ceiling assertion and the noise estimators.
\end{itemize}

\bibliographystyle{plainnat}
\bibliography{refs}

\appendix
\section{The BFCL Hard Stratum, Read Query by Query}\label{app:bfcl}

\emph{Section~\ref{sec:bfclcauses} states the three figures this appendix supports. What
follows is the full classification, the two families excluded from the stratified rows, and a
correction to our own scorer found while making it.}

\subsection{Why the hard stratum is hard}

Of the 121 held-out queries neither model answers from the shipped schema, \textbf{routing is
correct on 119}, so almost none of the failure is tool selection. \textbf{On 78 of the 121
both models emit a byte-identical call} that the key rejects, which is two independently
sampled models agreeing on a reading of the same text.

Classifying each by whether a rewrite could reach it: \textbf{111 of the 121 are
in-schema}, meaning every parameter the key wants is declared and the failure is about what
the description says about it. \textbf{Eight are
out-of-schema}, and they are the interesting ones. Seven sit in
\texttt{add\_postgres\_server}, whose answer key requires a \texttt{nickname} argument that
\textbf{the published schema never declares},
while the user's prompt supplies the value in seven of eight held-out queries. Neither model
answers any of them. No arm in this paper can: the parameter set is frozen for every method
(Section~\ref{sec:dataset}), so the call the key wants cannot be emitted at all. That is a
documentation defect of the severest kind occurring in a third-party benchmark, and it is
outside the action space we chose. The eighth sits in \texttt{get\_service\_providers}; both
families are excluded from the stratified rows.

\subsection{What blocks each call}

Counting in-schema failures says a rewrite is not structurally excluded. It does not say a
rewrite would help. We therefore read all 121 and assigned each a cause, then wrote each
assignment as a rule so the count reproduces. Table~\ref{tab:bfclcauses} is the result.

\begin{table}[t]
\centering
\small
\begin{tabular}{@{}lr@{}}
\toprule
\textbf{Cause} & \textbf{Queries} \\
\midrule
\multicolumn{2}{@{}l}{\emph{Reachable by rewriting the description}} \\
\quad Sentinel conflicts with the declared type & 50 \\
\quad Optional parameter volunteered unasked & 20 \\
\quad Value form not stated & 2 \\
\quad \textbf{subtotal} & \textbf{72 (59.5\%)} \\
\midrule
\multicolumn{2}{@{}l}{\emph{Not reachable that way}} \\
\quad Date with no year or reference date in the request & 10 \\
\quad Key mixes bare and qualified entity forms & 10 \\
\quad Key wants a value the request never states & 8 \\
\quad Key needs a parameter the schema omits & 7 \\
\quad Date stated but computed wrongly & 4 \\
\quad Key wants a value outside the schema's own enum & 4 \\
\quad Other value disagreement & 4 \\
\quad Wrong function & 2 \\
\quad \textbf{subtotal} & \textbf{49 (40.5\%)} \\
\bottomrule
\end{tabular}
\caption{What blocks each of the 121 held-out queries neither model answers from the shipped
schema. The division is whether the fix lies in the description of the function under test,
which is the action space every method here is confined to.}
\label{tab:bfclcauses}
\end{table}

\textbf{The largest single cause is a documentation defect of the kind our own dataset
plants deliberately.} In 50 of 121 the description tells the caller to pass a sentinel that
the declared type will not carry. \texttt{Music\_3\_LookupMusic.year} is declared
\texttt{integer} and its description reads ``Use 'dontcare' to include songs from any
year''. The \texttt{avg\_rating} of \texttt{get\_\allowbreak service\_\allowbreak providers}
is a \texttt{float} whose description says ``Use 'null' for unrated''. The
\texttt{smoking\_\allowbreak allowed} of \texttt{Hotels\_4\_\allowbreak SearchHotel} is
declared \texttt{string} while its key holds booleans.

\emph{The advice cannot be followed through a typed function-tool interface, and it is worth
being precise about what that does and does not mean.} Across every call either model
emitted, a sentinel string appears 721 times in 3{,}577 string-typed values and \textbf{zero
times} in 789 integer, 164 float and 944 boolean ones: the models reach for the sentinel
constantly and the interface refuses it wherever the type forbids it. \textbf{It does not
follow that the query is unanswerable.} Wherever the key names a sentinel the type forbids it
\emph{also} accepts the parameter being absent, and we checked that a call which simply omits
the parameter scores as correct in every such case. Across the whole held-out split there is
no query without a legal call.

So the defect is not a broken benchmark, and these queries should not be discarded. It is a
description that recommends an action the interface rejects while never mentioning the one
that works, which is a documentation failure of exactly the kind this paper measures.

Section~\ref{sec:bfclcauses} gives the argument this supports. Two details belong here
rather than there. The loop has a correctness signal the caller does not, because a
candidate rewrite is adopted only if it beats the incumbent on held-back training queries,
so a rewrite saying to omit the parameter when the request names no year is both derivable
from the schema alone and testable before it ships. And the contradiction is why the models
fail consistently rather than randomly: told to pass \texttt{dontcare} into an integer,
they invent a year or send 0 instead of leaving the parameter out.

\textbf{A further 20 are an optional parameter supplied unasked.} In the provider-search
family, ten queries ask only for a therapist in a named city. Both models add an
\texttt{insurance\_\allowbreak accepted} argument set true. The key lists no such argument,
so the call fails on the extra. Nothing in the description says when to leave a parameter
out.

\textbf{Against that, 49 are outside what any rewrite here can do.} Ten are dates the request
never grounds: ``find a Music event on March 14th'' with no year anywhere, scored against a
2023 key. Ten more are a key that spells the same kind of entity two ways inside one family,
wanting \texttt{Vancouver} in one query and \texttt{Sydney, AU} in another, so no single
stated convention satisfies both. Eight want a value the request never states, four want a
value the schema's own enum excludes, and seven need the parameter that is not declared.

\textbf{The arms fix the class we predicted and little else.} Of the queries \arm{audit}
rescued, four of six fall in the sentinel class and \arm{schedule} three of seven, against a
prior of 41\% if rescues were spread evenly. Four in fifty is also the rate at which the
class is repaired, so the reading is that the defect is real, reachable in principle and
mostly missed in practice. We draw no effect size from counts this small.

\subsection{A correction to our own scorer}

An earlier version of this comparison tested an emitted value against the key by type as
well as by value, so an integer 2019 failed a key entry of \texttt{"2019"} although the
schema declares that parameter an integer. Fifteen calls were scored wrong for obeying the
schema. The comparison now coerces across string, number and boolean spellings, every stored
run was re-scored offline from the recorded calls, and the baseline moves from 55.1\% to
59.1\%. We report it because it is the same class of defect this paper is about, in our own
instrument: a convention that one side of an interface states and the other does not honour.

\end{document}